\documentclass{iaus}
\usepackage{graphics}

  \checkfont{eurm10}
  \iffontfound
    \IfFileExists{upmath.sty}
      {\typeout{^^JFound AMS Euler Roman fonts on the system,
                   using the 'upmath' package.^^J}%
       \usepackage{upmath}}
      {\typeout{^^JFound AMS Euler Roman fonts on the system, but you
                   dont seem to have the}%
       \typeout{'upmath' package installed. iaus.cls can take advantage
                 of these fonts,^^Jif you use 'upmath' package.^^J}%
      }
  \else
  \fi

  \checkfont{msam10}
  \iffontfound
    \IfFileExists{amssymb.sty}
      {\typeout{^^JFound AMS Symbol fonts on the system, using the
                'amssymb' package.^^J}%
       \usepackage{amssymb}%

      }{}
  \fi

  \IfFileExists{amsbsy.sty}
    {\typeout{^^JFound the 'amsbsy' package on the system, using it.^^J}%
     \usepackage{amsbsy}}
    {\providecommand\boldsymbol[1]{\mbox{\boldmath $##1$}}}

\newsavebox{\astrutbox}
\sbox{\astrutbox}{\rule[-5pt]{0pt}{20pt}}

\title[Outskirts of Galaxy Clusters: intense life in the suburbs]
      {On the ellipticity of galaxy clusters}

\author[P. Flin {\it et al.\/}]%
{P. Flin, J. Krywult, M. Biernacka}

\affiliation{Institute of Physics, Pedagogical University, Kielce,
Poland}

\pubyear{2004}
\volume{195}
\pagerange{1--8}
\date{?? and in revised form ??}
\jname{Outskirts of Galaxy Clusters: intense life in the suburbs}
\editors{A. Diaferio, ed.}
\begin{document}

\maketitle

\begin{abstract}
We analysed 246 scanned ACO clusters and we do not find any
dependence of galaxy cluster ellipticity on redshift.
\end{abstract}

\section{Introduction}
The recent substantial evolution in the ellipticity of rich galaxy
clusters was noted by Mellot, Chambers \& Miller (2001). This
result has been supported by Plionis (2002). In particular, he
argues that the  ellipticity of  ACO clusters is correlated with
their redshifts. Using, among others, 309 ACO clusters taken from
APM, for which about 180 have determined redshifts, he claimed the
existence of a significant ellipticity - redshift dependence.

\section{Data}\label{sec:dat}
Our sample consists of 246 ACO clusters. All data come from 48''
Schmidt telescopes. Original plates were scanned at the Rome and
Edinburgh Astronomical Observatories. The majority of galaxy
catalogues were obtained from DSS using FOCAS packages. For this
sample, visual verification of object classification was made.

\section{Method of analysis}\label{sec:method}
It is well known that ellipticity depends also on the distance
from cluster centre (e.g. Carter \& Metcalfe 1980, Flin 1984,
Trevese et al. 1992, Struble \& Ftaclas 1994). Therefore, we
decided to check the investigated relation for various distances
from cluster centres. The radius changes from 0.5 to 1.5 Mpc ($h
=  0.75, q_0  = 0.5$) with a step of 0.25 Mpc from the cluster
centre. Thus, we have measured ellipticities at 5 distances from
the cluster centre for each of the 246  clusters. The
ellipticities have been determined in a standard way, this is from
the covariance matrix. The redshifts are taken from literature.

\begin{figure}
\centering
 \includegraphics{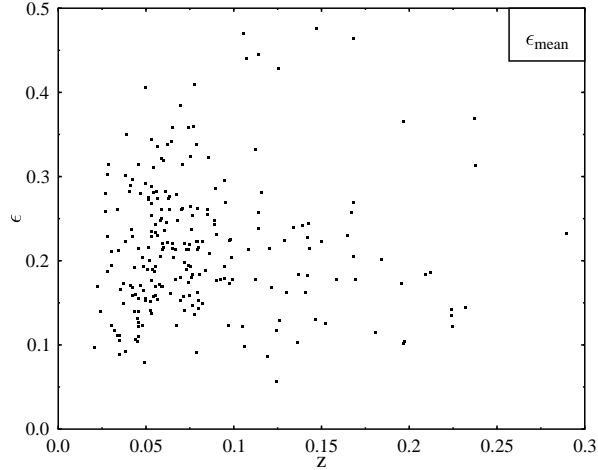}
  \caption{The relation between ellipticity and redshift for 246 investigated ACO clusters}\label{fig3}
\end{figure}

\section{Results}\label{sec:result}
For each cluster we have two parameters, the calculated
ellipticity of the cluster and cluster redshift. A straight line
was fitted to each set of data. The results of the fitting
procedure for line $e = A+B \cdot z$ are given in Table 1.
Parameter \emph{R} in Mpc is the radius of the area in which the
ellipse has been calculated, \emph{A} and  \emph{B} are fit
parameters and \emph{r} is the correlation coefficient. We also
calculated the mean value of the five ellipticities for various
\emph{R} radii. The dependence of such mean ellipticities on
redshift is given in Fig.1. The last line of Table 1 gives the
relevant parameters.

\section{Conclusions}\label{sec:concl}

It is clear that we do not observe any relation between
ellipticities and redshifts. This differs greatly from the Plionis
(2002) result. However, the claimed dependence of ellipticity on
redshift reported by him is very weak. Our result is affected by
such factors as e.g. the adopted cluster centre, the way of
smoothing the distribution and the applied method of ellipticity
determination. These factors are being carefully studied now. The
current result is consistent with Mulchaey \& Zabludoff (1998) and
Zabludoff \& Mulchaey (1998) observation reporting the lack of
recent significant dynamical evolution for poor groups and cD
clusters. Moreover, the lack of the eccentricity evolution is in
agreement with the latest hydrodynamic simulations of clusters
(Floor et al. 2003a, 2003b).

\begin{table}
\caption{The result of correlation analysis of the investigated
samples}
$$\begin{tabular}{cccccc}
 \hline \hline
\emph{N} &\emph{R} & \emph{A} & \emph{B} & \emph{r} \\
\hline \hline
  1 & 0.50 & 0.298 & 0.218 & 0.05 \\
  2 & 0.75 & 0.215 & 0.118 & 0.05 \\
  3 & 1.00 & 0.207 & -0.097 & 0.05 \\
  4 & 1.25 & 0.186 & -0.115 & 0.06 \\
  5 & 1.50 & 0.176 & -0.061 & 0.04 \\
\hline \hline
  6 & Mean & 0.216 & 0.013 & 0.01 \\
\hline \hline
\end{tabular}$$
\end{table}

\bigskip
\bigskip
\bigskip

\begin{acknowledgments}
This paper was financially supported by Pedagogical University
grants: BS 052, BW 116 and BW 236.
\end{acknowledgments}


\begin{thebibliography}{}

\bibitem [Carter \& Metcalfe (1980)]{Carter80}{Carter, D., Metcalfe, N.} 1980 MNRAS, 191, 325
\bibitem [Flin (1984)]{Flin84}{Flin, P.} 1984
in Clusters and Groups of Galaxies, eds. Mardirossian, F., et al., Dordrecht, p. 163
\bibitem [Floor, Melott, Miller, and Bryan (2003)]{Floor03} {Floor, S.N., Melott, A.L., Miller, C.J., Bryan, G.L.} 2003a, ApJ, 591, 741
\bibitem [Floor, Melott and Motl(2003)]{Floor03}{Floor, S.N., Melott, A.L., Motl,
P.M.} 2003b, astro-ph/0307539
\bibitem [Melott, Chambers, Miller]{Melott01} {Melott, A., Chambers, S. W., Miller, C. J.} 2001, ApJ, 559, L75
\bibitem [Mulchaey \& Zabludoff (1998)]{Mulchaey98} {Mulchaey, J. S., Zabludoff, A. I.} 1998, ApJ, 496, 73
\bibitem [Plionis (2002)]{Plionis02} {Plionis, M.} 2002, ApJL, 572, 67
\bibitem [Struble \& Flaclas (1994)]{Struble94} {Struble, M. C., Flaclas, F.} 1994, AJ, 108, 1
\bibitem [Trevese, Flin, Migliori, Hickson, Pittella (1992)]{Trevese92}
{Trevese, D., Flin, P., Migliori, L., Hickson, P., Pittella, G.}
1992, A\&AS, 94, 237
\bibitem [Zabludoff \& Mulchaey (1998)]{Zabludoff98} {Zabludoff, A. I., Mulchaey, J. S.} 1998, ApJ, 496, 39

\end{thebibliography}
\end{document}